\documentclass[10pt]{article}
\usepackage{amssymb}
\usepackage{amsthm}
\usepackage{amsmath}
\usepackage[noblocks]{authblk}
\usepackage[a4paper, total={6.5in, 10in}]{geometry}
\usepackage{braket}
\usepackage[version=3]{mhchem}
\usepackage{pst-3dplot}
\usepackage{appendix}

\newtheorem{theorem}{Theorem}
\newtheorem*{remark}{Remark}
\newtheorem{definition}{Definition}

\newtheorem{assumption}{Assumption}

\begin{document}
\title{Pseudo-mass parameterized alchemical equation: a generalisation of the molecular Schr\"odinger equation}
\author[1]{Qing-Long Liu}
\affil[1]{Department of Chemistry,  University of Basel,  Klingelbergstrasse 80, CH-4056 Basel, Switzerland}
\affil[1]{\textit {qinglong.liu@unibas.ch}}
\maketitle

\begin{abstract}
We introduce a pseudo-mass parameterized Schr\"odinger-like alchemical equation which contains nuclear charges as variables, treating nuclear charges, nuclear coordinates and electronic coordinates on the equal footing. The eigenfunctions of the alchemical equation are the wave functions of nuclear charges, just like conventional wave functions are of coordinates. A mathematical definition of alchemical function space is given to hold the ``nuclear charge wave function''. The geometric phase of alchemical dynamics and alchemical phase space are also derived. For hydrogen-like ion, the alchemical equation can be simplified into a strong repulsive inverse square potential equation, which refers to the quantum anomaly and keeps conformal invariance in non-relativistic quantum mechanics. An extension of the Hellmann-Feynman theorem, which applies to the non-stationary state in time-dependent clamped-nuclear-charge alchemical dynamics, is also proved.
\end{abstract}

\section{Introduction} \label{sec:1}
As a matter of fact, nuclear charge is a positive integer number and used as a fixed parameter in Schr\"odinger equation. 
In chemistry study, nuclear charge sometimes is computed continuously with derivative and called alchemical derivative \cite{VonLilienfeld2005,Balawender2019}. 
From the thermodynamics point of view, the energy change during a chemical reaction is path independent. It could go either along a coordinate path or an alchemical path or both. 
For instance, a chemical reaction \ce{Cl2 + H2 = 2 HCl}. Both \ce{Cl-Cl} and \ce{H-H} bonds break, chlorine atoms and hydrogen atoms are rearranged in space during the reaction. Alternatively, we could consider one chlorine atom goes along an alchemical path and becomes a hydrogen atom. One hydrogen atom goes backward to the chlorine atom. Next, the \ce{H-Cl} bond length adjusts a little bit to reach a stable structure. 
These two ways are equal in thermodynamics sense guaranteed by Hess's law \cite{Chang2014,Hahn2019}. 
In this article, we directly take nuclear charges as continuous variables in Schr\"odinger equation from the computational point of view and add a new kinetic energy term $-\frac{1}{\mu} \Delta_{Z}$ for nuclear charge $Z$. Here $\mu$ is the ``pseudo-mass'' of nuclear charge, and it will alter the kinetic properties of alchemical dynamics. Also, nuclear charge is a positive real number instead of a positive integer.

Table \ref{diffdyna} lists a comparison between different quantum level dynamics methods, so classical molecular dynamics, single point energy calculation, etc are not included. Ab-initio molecular dynamics includes Born-Oppenheimer molecular dynamics, Car–Parrinello molecular dynamics, etc \cite{Marx2009}. In this class of methods, nuclear coordinate $R$ and nuclear charge $Z$ keep as fixed parameters in the corresponding equations. The force applying on the nucleus comes from a quantum level calculation, but the nuclear motion obeys classical Newtonian equation, and moves along nuclear coordinate $R$. This method could be seen as a clamped-nucleus quantum molecular dynamics. Quantum molecular dynamics is a broad research field. It has many different variants \cite{Gatti2014}. In general, both nucleus and electron move quantum mechanically along the nuclear coordinate $R$. In this article, we will discuss alchemical dynamics that both nuclear coordinates $R$ and nuclear charge $Z$ are variables. This class of methods describes the dynamics behaviour of both nuclear coordinate $R$ and nuclear charge $Z$. We could also set nuclear charge as a fixed parameter in alchemical dynamics and still move along $R$ and $Z$, this could be seen as a clamped-nuclear-charge alchemical dynamics. The remaining two types of dynamics in Table \ref{diffdyna} are in similar manners, but we will not go there in this article. 

Before giving the alchemical equation, we first define the exact basis function for the alchemical equation. Next, we derive the general alchemical equation with an unknown pseudo-mass in Sec.~\ref{sec:2}. Sec.~\ref{sec:3} introduce the definition of alchemical function space which is an abstract structure for the eigenfunctions of the alchemical operator. The geometric phase of the nuclear charge wave function is derived in Sec.~\ref{sec:4}. A new ``alchemical momentum'' operator correspondence is defined on $\mathbb{R}_{> 0}$, and further, the alchemical phase space is given in Sec.~\ref{sec:7}. In Sec.~\ref{sec:5}, we consider the simplest case of the alchemical equation, i.e., hydrogen-like ion, a system with varying nuclear charge and only one electron. 
We obtain a strong repulsive inverse square potential equation, make it as another example of the inverse square potential application. Finally, in Sec.~\ref{sec:6}, 
we derive the expression of extended Hellmann-Feynman force of nuclear charge
for the non-stationary state in clamped-nuclear-charge alchemical dynamics. All the proofs are given in the Appendix. 

\begin{table}
\caption{Comparing different quantum level dynamics} \label{diffdyna}
\centering
\begin{tabular}{|c|c|c|c| } 
 \hline
 Types of the dynamics & Coordinate $R$ & Charge $Z$ & Dynamics of \\ \hline
Ab-initio molecular dynamics  & parameter & parameter & $R$\\ 
Quantum molecular dynamics  &  variable & parameter & $R$ \\
- &  parameter & variable & $R$ \\
- &  parameter & variable & $Z$ and $R$ \\
Clamped-nuclear-charge alchemical dynamics & variable & parameter & $Z$ and $R$ \\
Alchemical dynamics & variable & variable & $Z$ and  $R$ \\ \hline
\end{tabular}
\end{table}


\section{Pseudo-mass parameterized alchemical equation} \label{sec:2}
\begin{assumption}[Nuclear charge kinetic term and large pseudo-mass] \label{ass:1}
We assume that there is a kinetic term $-\frac{1}{\mu} \Delta_{Z}$ of nuclear charge $Z$ in alchemical operator
\begin{equation}
    \hat{H} = \sum_A \left( -\frac{1}{\mu} \Delta_{Z_A} - \frac{1}{m_A} \Delta_{R_{A}} + H_e + \frac{1}{2}\sum_{B \neq A} \frac{Z_A Z_B}{|R_A-R_B|} \right)  \label{eq:27}
\end{equation}
where $\mu$ is a constant pseudo-mass for any atom and any nuclear charge. $\mu \gg m$ where $m$ is atomic mass and nuclear charge $Z \in \mathbb{R}_{\geqslant 0}$. $H_e$ is the electronic Hamiltonian without nuclear repulsion term.
\end{assumption}
A potential issue of this operator is that the self-adjoint property might not hold. For example, operator $\Delta_{Z}$ defined on the domain $0 \le Z < \infty $ is not self-adjoint \cite{Bonneau2001}. It's deficiency indices is $(1,1)$, and therefore it has infinitely many self-adjoint extensions. One could solve it by setting boundary condition $\psi(Z=0)=0$ where $\psi$ is the eigenfunction of this operator \cite{Araujo2004}. The self-adjoint property also depends on the potential term, as we will see in hydrogen-like ion case in Sec.~\ref{sec:5}, the pseudo-mass parameterized alchemical equation can be simplified into inverse square potential Hamiltonian  (\ref{eq:3}). And further it could be renormalized and recover its self-adjoint character \cite{Scursulim2020}. 

\begin{definition}[Exact wave function]
Based on assumption \ref{ass:1}, the exact wave function of alchemical operator is as follows
\begin{align}
\Psi(r,R, Z) &  = \sum_{n,j,k}^{+\infty} c_{n,j,k}\ \eta_{n,j,k}(Z) \otimes \chi_{j,k}(R | Z) \otimes \phi_k(r | (R, Z))  \label{eq:28} \\
&\approx \sum_{n,j}^{+\infty} c_{n,j}\ \eta_{n,j}(Z) \otimes \chi_j(R | Z) \otimes \phi_j(r | (R, Z)) \label{eq:11} \\
& \approx \sum_n^{+\infty} \eta_n(Z) \otimes \chi_n(R | Z) \otimes  \phi_n(r | (R, Z))\label{eq:12} 
\end{align}
and satisfy orthonormality constraints
\begin{equation}
 \begin{gathered}
     \Braket{\Psi_i | \Psi_{i'}} = \delta_{i,i'} \qquad \Braket{\eta_i | \eta_{i'} } = \delta_{i,i'}  \\
     \Braket{\chi_i | \chi_{i'}} = \delta_{i,i'} \quad \forall Z \qquad  \Braket{\phi_i | \phi_{i'}} = \delta_{i,i'} \quad \forall R, Z  \label{eq:26}
 \end{gathered}
\end{equation}
where $\eta, \chi, \phi$ are the wave functions of nuclear charge $Z \in \mathbb{R}_{\ge 0}$,
nuclear and electronic coordinates $R\in \mathbb{R}$, $r\in \mathbb{R}$, respectively.
$\phi(r | (R, Z))$ means electronic wave function $\phi(r)$ at a fixed coordinate $R$ 
and a fixed nuclear charge $Z$.
$\chi(R | Z)$ means molecular wave function $\chi(R)$ at a fixed nuclear charge $Z$.
\end{definition}

There are two ways of writing approximate wave functions, expression (\ref{eq:11}) and expression (\ref{eq:12}).
Expression (\ref{eq:11}) is based on the Born-Oppenheimer approximation only. Electronic wave function and molecular wave function do not entangle with each other since a nucleus is much heavier than an electron. Based on assumption \ref{ass:1}, ``nuclear charge wave function'' $\eta(Z)$ do not entangle with other parts as well, that is expression (\ref{eq:12}). This is a reasonable assumption because the nuclear charge does not change in reality, and it could be approximately treated as it is much more heavier in dynamics. In physics literature, tensor product $\otimes$ in expressions (\ref{eq:28}), (\ref{eq:11}) and (\ref{eq:12}) are often omitted without any ambiguity. 

$\chi (R|Z)$ is the eigenfunction of the molecular Hamiltonian, it satisfy
\begin{equation}
    \left (-\sum_A \frac{1}{m_A} \Delta_{R_A} + \Bra{\phi} H_e \Ket{\phi} \right ) \chi  =  E \chi \label{eq:13}
\end{equation}

$\phi(r|(R,Z))$ is the eigenfunction of the electronic Hamiltonian $H_e$, it satisfy
\begin{equation}
     \left (-\sum_i\Delta_{r_i} - \sum_i \sum_{A} \frac{Z_A}{|r_i-R_A|} + \sum_{i > j} \frac{1}{|r_i-r_j| } \right ) \phi = E \phi \label{eq:14}
\end{equation}

What we need now is the operator for $\eta(Z)$. Based on the wave function defined in (\ref{eq:12}), the definition of the pseudo-mass parameterized alchemical equation is as follows.
\begin{theorem}[Pseudo-mass parameterized alchemical equation] \label{the:1}
The pseudo-mass parameterized alchemical equation is defined as
\begin{equation}
    \begin{gathered}
         \sum_{A} \bigg(  -\frac{1}{\mu} \Big(\Delta_{Z_A} + \Bra{\chi}\Delta_{Z_A}\Ket{\chi} + \Bra{\phi} \Delta_{Z_A} \Ket{\phi} \Big)
         - \frac{1}{m_A} \Big(\Bra{\chi}\Delta_{R_{A}}\Ket{\chi} +\Bra{\phi}\Delta_{R_{A}}\Ket{\phi} \Big) \\
         +\Bra{\phi} H_e \Ket{\phi} + \Bra{\chi} \frac{1}{2}\sum_{B \neq A} \frac{Z_A Z_B}{|R_A-R_B|} \Ket{\chi} \bigg) \eta =E \eta  \label{eq:1}
    \end{gathered}
\end{equation}
where $\eta, \chi, \phi$ are wave functions of nuclear charge $Z \in \mathbb{R}_{\geqslant 0}$,
nuclear and electronic coordinates $R\in \mathbb{R}$, $r\in \mathbb{R}$, respectively.
$\mu$ is the pseudo-mass for nuclear charge and $m$ is the atomic mass.
\end{theorem}

The proof of this theorem is in \ref{app:1}. Solving Eq.~(\ref{eq:1}) is a high dimensional problem. We need an ``alchemical potential energy surface'' which refers to the total energy function of different spatial structures and nuclear charges of chemical species. The problem is the pseudo-mass $\mu$ and continuous atomic mass function $m(Z)$ are not given. Only $m(Z)$ of integer $Z$ are known. One could choose a reasonable large $\mu$ to perform a pseudo-dynamics simulation to represent a chemical reaction, as mentioned in the introduction part. However, the kinetic information coming out of pseudo-dynamics does not correspond to a real reaction kinetics. If $\mu$ is small and there will be a coupling between nuclear charge dynamics and molecular dynamics, then the approximation of exact wave function in Eq.~(\ref{eq:12}) is no longer valid. 
In Sec.~\ref{sec:5} we will see that the pseudo-mass problem could be bypassed in hydrogen-like ion case by introducing the pseudo-mass weighted nuclear charge.


\section{Alchemical function space  $\mathbb{V}_{ACS}$}  \label{sec:3}
\begin{definition}[Alchemical function space] \label{def:2}
The alchemical function space $\mathbb{V}_{ACS}$ has a fiber bundle structure which consists of three spaces. 
\begin{equation}
	\mathbb{V}_{ACS} :=\mathbb{V}_{Z}\times\mathbb{V}_{R|Z}\times\mathbb{V}_{e|(R,Z)} \label{eq:24}  
\end{equation}
with following conditions hold
\begin{equation*}
\begin{split}
	\text{fiber bundle map } f_1 & : \mathbb{V}_{ACS} \rightarrow \mathbb{V}_{Z} \\
	\text{fiber bundle map } f_2 & : \mathbb{V}_{R|Z} \times \mathbb{V}_{e|(R,Z)} \rightarrow \mathbb{V}_{R|Z}
\end{split}
\end{equation*}
$\mathbb{V}_{Z}$
is a direct sum of infinite quotient spaces.
\begin{equation}
\mathbb{V}_{Z}= \bigoplus_{N_n=0}^{+\infty} \mathbb{V}_{Z,1}^{\otimes N_n} \label{eq:25}   
\end{equation}
where $\mathbb{V}_{Z,1}=L^2(\mathbb{R}_{>0})$ is for holding the nuclear charge wave function of 
one nucleus.
At a given number of atoms $N_n$, 
\begin{equation}
\mathbb{V}_{Z|N_n}=  L^2(\mathbb{R}_{\ge 0}^{N_n}) \label{eq:22}  
\end{equation}
\end{definition}

The operator $\times$ is Cartesian product,
$\mathbb{V}_{e|(R,Z)}= \bigoplus_{N_e=0}^{+\infty} H^{\wedge {N_e}}  $ is the Fock space of electronic states, $H=L^2(\mathbb{R}^3)$. At a given number of electrons $N_e$, $\mathbb{V}_{e|(R,Z,N_e)} = L^2(\mathbb{R}^{3 N_e})$. $N_e$ is the number of electrons in the system.
$\mathbb{V}_{R|Z}=L^2(\mathbb{R}^{3N_n-6(5)})$ is the Hilbert space of molecular wave functions at a given $Z$.  $N_n$ is the number of atoms in the system.
$\mathbb{V}_{Z}$ is the space of nuclear charge wave functions. 
$\mathbb{V}_{ACS}$ is the total space in $f_1$ fiber bundle, its base space is $\mathbb{V}_{Z}$.
$\mathbb{V}_{R|Z} \times \mathbb{V}_{e|(R,Z)}$ is the total space in $f_2$ fiber bundle, its base space is $\mathbb{V}_{R|Z}$. 

\begin{remark}
A symmetric group $S_{N_n}$ should apply on elements in $\mathbb{V}_{Z}$ and $\mathbb{V}_{Z|N_n}$ in order to remove duplicate elements in alchemical function space since the sequence of elements of a chemical compound is not essential.
For instance,
$\eta(1, 6, 7)$ and $\eta(6, 7, 1)$ are identical, and both of them represent the nuclear charge wave function of hydrogen cyanide molecule \ce{HCN}, see figure \ref{P1}.
\end{remark}

The number of atoms $N_n$ does not necessarily conserve during an alchemical reaction
if we take $Z \in \mathbb{R}_{>0}$.
It could range from $N_n$ to $2N_n-1$ depending on the change of nuclear coordinates.
If all products occupy the same spatial coordinates as reactants, only the type of atoms
are exchanged after the chemical reaction. For such cases, we consider the number of atoms keeps conservative during the whole alchemical reaction path. 
However, if every atom changes its coordinates after the reaction, the number of 
atoms could be up to $2N_n-1$ during the alchemical reaction. 
We continue to use the example in the introduction part,
a chemical reaction \ce{Cl2 + H2 = 2 HCl}. 
If \ce{Cl-Cl}, \ce{H-H} and \ce{H-Cl} bonds have the same lengths and same positions, 
then the number of atoms does not change during the alchemical reaction.
If every atom change its coordinates after the reaction,
we could consider reactants' nuclear charges go to zero (except the one used as origin point), 
some new atoms emerged in space and nuclear charges go from zero to $1$ (\ce{H}) or $17$ (\ce{Cl}).
During this process, there are at most $7$ atoms along the alchemical path. 
The number of atoms $N_n$ does keep conservative during an alchemical reaction
if we take $Z \in \mathbb{R}_{\ge 0}$ and $N_n$ is large enough
since $Z = 0$ preserves some auxiliary atoms.

Table \ref{compare} gives comparisons of wave functions and spaces in the static case,
quantum dynamics case and alchemical case, respectively.
The nuclear charge part, or saying
$\mathbb{V}_{Z}$, is the new space for the nuclear charge wave functions.

\begin{table} 
\caption{wave functions and Spaces' Comparisons} \label{compare}
\centering
\begin{tabular}{|c|c|c|c|}
\hline
  & Statics & Dynamics & Alchemy  \\ \hline
 wave function & $\phi(r)$ & $\chi(R)\otimes\phi(r|R)$ & $\eta(Z)\otimes\chi(R|Z)\otimes\phi(r|(R,Z))$ \\ 
 Space & $\mathbb{V}_e$ & $\mathbb{V}_{R}\times\mathbb{V}_{e|R}$ & $\mathbb{V}_{Z}\times\mathbb{V}_{R|Z}\times\mathbb{V}_{e|(R,Z)}$ \\     \hline
\end{tabular}
\end{table}

\psset{unit=1.0cm, coorType=6}
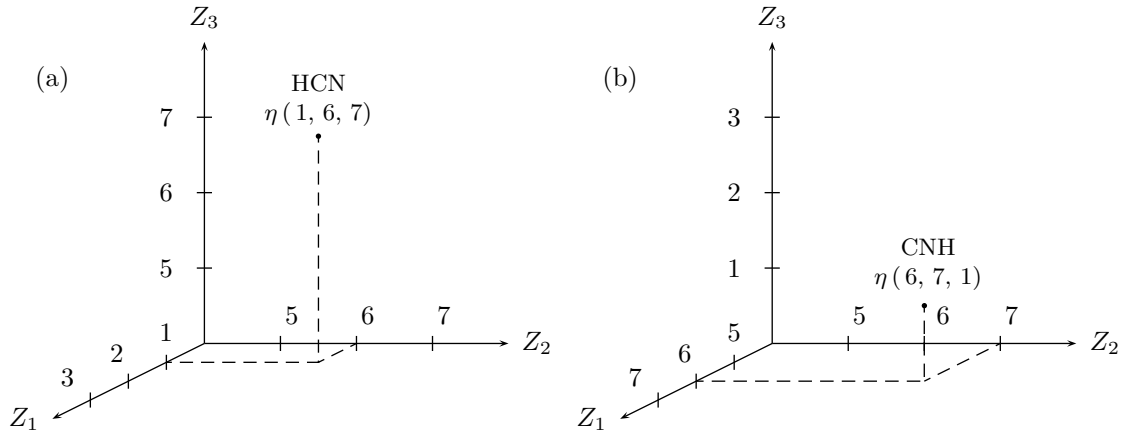
\begin{figure}
\centering
\begin{pspicture}(-2.5,-1.5)(4.5,4.5)
	\psset{%
		IIIDxTicksPlane=xz,
		IIIDyTicksPlane=yz}
	\pstThreeDCoor[%
		xMin=0,xMax=4, 
		yMin=0,yMax=4,
		zMin=0,zMax=4,
		IIIDticks,
		spotX=180,
		nameX=$Z_{1}$,
		nameY=$Z_{2}$,
		nameZ=$Z_{3}$,
		IIIDlabels=false,
		linecolor=black]%
	\pstThreeDPut(1.0,2.0,3.3){\small $\eta\,(\,1,\,6,\,7)$}
	\pstThreeDPut(1.0,2.0,3.7){\small \ce{HCN}}
	\pstThreeDPut(2,-1,4){(a)}
	\psset{dotstyle=*,dotsize=2pt,drawCoor=true}
	\pstThreeDDot(1,2,3)
		\pstThreeDPut(3.2,-0.2,0.35){$3$}%
		\pstThreeDPut(1.9,-0.2,0.35){$2$}%
    	\pstThreeDPut(1.0,-0.,0.4){$1$}%
		\pstThreeDPut(-0.3,1,0.3){$5$}%
		\pstThreeDPut(-0.3,2,0.3){$6$}%
		\pstThreeDPut(-0.3,3,0.3){$7$}%
		\pstThreeDPut(0,-0.5,1){$5$}
		\pstThreeDPut(0,-0.5,2){$6$}
		\pstThreeDPut(0,-0.5,3){$7$}
\end{pspicture}
\quad
\begin{pspicture}(-2.5,-1.5)(4.5,4.5)
	\psset{%
		IIIDxTicksPlane=xz,
		IIIDyTicksPlane=yz}
	\pstThreeDCoor[%
		xMin=0,xMax=4, 
		yMin=0,yMax=4,
		zMin=0,zMax=4,
		IIIDticks,
		spotX=180,
		nameX=$Z_{1}$,
		nameY=$Z_{2}$,
		nameZ=$Z_{3}$,
		IIIDlabels=false,
		linecolor=black]%
	\pstThreeDPut(2.1,3.1,1.4){\small $\eta\,(\,6,\,7,\,1)$}
	\pstThreeDPut(2.1,3.1,1.8){\small \ce{CNH}}
	\pstThreeDPut(2,-1,4){(b)}
	\psset{dotstyle=*,dotsize=2pt,drawCoor=true}
	\pstThreeDDot(2,3,1)
		\pstThreeDPut(3.2,-0.2,0.35){$7$}%
		\pstThreeDPut(1.9,-0.2,0.35){$6$}%
    	\pstThreeDPut(1.0,-0.,0.4){$5$}%
		\pstThreeDPut(-0.3,1,0.3){$5$}%
		\pstThreeDPut(-0.3,2.1,0.3){$6$}%
		\pstThreeDPut(-0.3,3,0.3){$7$}%
		\pstThreeDPut(0,-0.5,1){$1$}
		\pstThreeDPut(0,-0.5,2){$2$}
		\pstThreeDPut(0,-0.5,3){$3$}
\end{pspicture}
\caption{Two identical nuclear charge wave functions (a) $\eta(1,6,7)$ and (b) $\eta(6,7,1)$ of hydrogen cyanide \ce{HCN} molecule used as elements in $\mathbb{V}_{Z|(N_n=3)}$ space.
Each dimension represents an atom and the values along each axis are nuclear charges.
\ce{HCN}, \ce{HNC} \ce{CNH}, \ce{CHN}, \ce{NCH} and \ce{NHC} are all belong to $S_3$ symmetric group. \label{P1}}
\end{figure}


\section{Geometric phase of alchemical equation} \label{sec:4}

When degeneracy of alchemical potential surface is included as system evolving adiabatically round a closed path in $R(t)$ or $Z(t)$, non-trivial geometrical phase factor $\exp(i \gamma(C))$ is emerged \cite{BERRY1984}. We now assume nuclear charge $Z(t)$ and molecular coordinate $R(t)$ evolve with time, but not electronic coordinate $r$, and alchemical operator (\ref{eq:27}) does not commute at different times
\begin{equation}
    [\hat{H}(t),\hat{H}(t')] \neq 0 \label{eq:7}
\end{equation}
Hence eigenfunctions $\Psi(t)$ of $\hat{H}(t)$ at different time are different.
For time $T$, we have
\begin{definition}[wave function with geometrical phase factor] \label{def:3}
The stationary time-dependent alchemical total wave function with geometrical phase factor is
\begin{equation}
    \begin{gathered}
    \Psi_n(t) =  \exp{\left(-\frac{i}{\hbar}\int_0^T E_n\big(Z(t'),R(t')\big) d t' \right)} \exp{(i\gamma_n(t))}~ \eta_n(Z(t)) \\
    \chi_n(R(t)|Z(t))~ \phi_n(r | (R(t), Z(t))) \label{eq:65}   
    \end{gathered}
\end{equation}
\end{definition}
which is based on Eq.~(\ref{eq:12}). $\gamma_n(C)$ is the geometrical phase  \cite{BERRY1984,Mead1992,Bohm1992}. $n$ represents different eigenfunctions. 

\begin{theorem}[Geometric phase] \label{the:2}
The geometric phase of time-dependent version alchemical equation is
        \begin{equation}
            \gamma_n(C)  = i \oint_{C_R} \Bra{\chi_n}\nabla_R \Ket{\chi_n} d R + i \oint_{C_Z} \Big( \Bra{\chi_n} \nabla_Z \Ket{\chi_n} + \Bra{\eta_n}\nabla_Z \Ket{\eta_n} \Big) d Z  \label{eq:4} 
        \end{equation}
$C_R$ and $C_Z$ are closed loops along nuclear coordinate path and nuclear charge path, respectively. 
\end{theorem}
The proof of this theorem is in \ref{app:2}. This geometric phase change has three parts, molecular wave function evolve along $R$, nuclear charge wave function evolve along $Z$ and an additional mixed term, $i \oint_{C_Z} \Bra{\chi_n} \nabla_Z \Ket{\chi_n} d Z$, molecular wave function evolve along $Z$.
Usually, conical intersection of potential energy surface comes out when the degree of freedom is larger and equal than 2. For a molecule with $N$ atoms, its degree of freedom is $3N-6(5)$ without the consideration of spin. Only $N \ge 3$ molecules have a conical intersection.
However, in the alchemical case, conical intersection comes into play when $N \ge 2$ since the degree of freedom in nuclear charge $Z$ parameter space is equal to the number of atoms $N$. 


\section{Alchemical phase space} \label{sec:7}
Momentum is not a measurable quantity if it is defined on positive semi-axis. Such momentum operator has deficiency indices $(1,0)$, which means it has no self-adjoint extension \cite{Bonneau2001}.  Therefore it is problematic to define a momentum-like operator $\hat{p}_Z = -i\hbar \nabla_Z$ on $\mathbb{R}_{\ge 0}$. 
However one could give close correspondences to the $\hat{x}$, $\hat{p}$ and make them satisfy Heisenberg algebra $[\hat{x},\hat{p}]=i\hbar$ relation \cite{Twamley2006}. 

\begin{definition}[Alchemical phase space] \label{the:5}
An alchemical phase space is denoted as pair $(\hat{\theta},\hat{p_{\theta}})$ such that
\begin{equation}
    \begin{gathered}
    \hat{\theta}  = \ln Z   \\ 
    \hat{p}_{\theta}  = \frac{1}{2} \left(\hat{Z}\hat{p}_Z+\hat{p}_Z\hat{Z}\right) = Z p_Z - \frac{i\hbar}{2} = -i\hbar \left(Z\nabla_Z+\frac{1}{2}\right) \label{eq:35} \end{gathered}
\end{equation}
where $\hat{p}_Z = -i\hbar \nabla_Z$. $\hat{\theta}$ is defined on $\mathbb{R}_{> 0}$ and $\hat{p}_{\theta}$ is essentially self-adjoint on $\mathbb{R}_{\ge 0}$ \cite{Twamley2006}. 
Moreover Heisenberg algebra 
\begin{equation}
    [\hat{\theta},\hat{p}_{\theta}] = i\hbar \label{eq:48}
\end{equation}
holds.
\end{definition}
To verify Eq.~(\ref{eq:48}) we have
\begin{align*}
    [\hat{\theta},\hat{p}_{\theta}]& = \hat{\theta}\hat{p}_{\theta}-\hat{p}_{\theta}\hat{\theta}\\
    & = \ln Z \left( -i\hbar \left(Z\nabla_Z+\frac{1}{2}\right) \right) -  \left( -i\hbar \left(Z\nabla_Z+\frac{1}{2}\right) \right) \ln Z \\
    & = -i\hbar \bigg( \ln Z  Z  \nabla_Z+\frac{1}{2}\ln Z - Z \ln Z \nabla_Z - 1 -\frac{1}{2}\ln Z     \bigg) \\
    & = i\hbar
\end{align*}
If we define $\hat{\theta} = Z$, then both $\theta$ and $\hat{p}_{\theta}$ could be defined on $\mathbb{R}_{\ge 0}$ and satisfy
\begin{align*}
    [\hat{\theta},\hat{p}_{\theta}]& = Z \left( -i\hbar \left(Z\nabla_Z+\frac{1}{2}\right) \right) -  \left( -i\hbar \left(Z\nabla_Z+\frac{1}{2}\right) \right)  Z \\
    & = -i\hbar Z^2 \nabla_Z - \frac{1}{2} i\hbar Z + i\hbar Z + i\hbar Z^2 \nabla_Z + \frac{1}{2} i\hbar Z \\
    & = i\hbar Z
\end{align*}

\section{Alchemical equation and inverse square potential} \label{sec:5}
In this section, we consider a simple example of the pseudo-mass parameterized alchemical equation.
\begin{theorem}[Equivalent to inverse square potential] \label{the:3}
For the ground state hydrogen-like ion, the alchemical equation (\ref{eq:1}) could be simplified to
        \begin{equation}
            \left( -\Delta_{Z_w} + \frac{0.75}{Z_w^2} -\frac{Z_w^2 Ry}{\mu} \right) \eta_w = E_w \eta_w \label{eq:2}
        \end{equation}
$Ry$ is the Rydberg energy. $Z_w$ is the pseudo-mass weighted nuclear charge, $Z_w = \sqrt{\mu} Z$.
$E_w$ and $\eta_w$ are the pseudo-mass weighted energy and the pseudo-mass weighted nuclear charge wave function, respectively.
We could further take $\mu \to +\infty$ and get
        \begin{equation}
            \left( -\Delta_{Z_w} + \frac{0.75}{Z_w^2} \right) \eta_w = E_w \eta_w \label{eq:3}
        \end{equation}
\end{theorem} 
The proof of this theorem is in \ref{app:3}. We notice that the Hamiltonian in Eq.~(\ref{eq:3}) is not self-adjoint and it has a strong repulsive inverse square potential which refers to the anomaly in quantum mechanics \cite{Coon2002,Essin2006} as well as non-relativistic conformal field theory \cite{Moroz2010,Nishida2007}.
Renormalization is one of the methods to deal with this potential \cite{Camblong2000} and 
it turns out that a self-adjoint extension is equivalent to a renormalization procedure \cite{Scursulim2020}. Similar analysis could further confirm the fact that strong repulsive inverse square potential also exist in first excited state, second excited state, etc. Our alchemical equation is based on the assumption that $\mu \gg m$. 
In reality, nuclear charge does not change hence we could further take $\mu \to +\infty$. 
The solution of Eq.~(\ref{eq:3}) corresponds to an extreme small displacement at $Z=0$.
If $\mu$ is in a similar magnitude as $m$ or even smaller, Eq.~(\ref{eq:12})
will no longer hold and one should use Eq.~(\ref{eq:11}) instead. If so, the fiber bundle structure
in definition \ref{def:2} will break as well.


\section{Hellmann-Feynman theorem extension for non-stationary state} \label{sec:6}
As shown in the Table \ref{diffdyna}, clamped-nuclear-charge alchemical dynamics is by setting
nuclear charges as a fixed parameters in alchemical dynamics but still move along $R$ and $Z$. 
\begin{definition}[Clamped-nuclear-charge alchemical dynamics] \label{def:5}
The clamped-nuclear-charge alchemical dynamics Hamiltonian and stationary clamped-nuclear-charge  wave function which commutes at different times are
\begin{align}
\hat{H}_{clamped}  & = \sum_A \left( -\frac{1}{m_A} \Delta_{R_A} + H_e +\frac{1}{2}\sum_{B\neq A} \frac{Z_A Z_B}{|R_A-R_B|} \right) \label{eq:33} \\
\Psi_{clamped,n} (t) &  = \exp{\left(-\frac{i}{\hbar}\int_0^T E_n\big(Z(t')\big) d t' \right)} ~ \chi_n(R|Z)~ \phi_n(r|(R,Z)) \label{eq:29}  
\end{align}
\end{definition}
we have
\begin{align}
    \nabla_Z E & = \nabla_Z \left( \Bra{\Psi_{clamped,n}(t)} \hat{H}_{clamped} \Ket{\Psi_{clamped,n}(t)} \right) \nonumber \\
    & = \Bra{\Psi_{clamped,n}(t)} \nabla_Z \hat{H}_{clamped} \Ket{\Psi_{clamped,n}(t)}  \label{eq:6}
\end{align}
which is conventional Hellmann-Feynman force on $Z$.
However, for non-stationary state in time-dependent clamped-nuclear-charge alchemical dynamics, 
Hellmann-Feynman theorem is no longer valid and needs an extension, 
just like in non-stationary molecular dynamics simulation \cite{Bala1994}.
\begin{definition}[Non-stationary clamped-nuclear-charge wave function] \label{def:6}
The non-stationary clamped-nuclear-charge wave functions are defined as
\begin{equation}
\Psi_{ns}(t)  = \sum_n c_n(t) \Psi_{clamped,n}(t)  \label{eq:23} \\ 
\end{equation}
where 
\begin{equation}
    \sum_n c_n(t)  = 1 \label{eq:32} 
\end{equation}
are the normalized coefficients.
\end{definition}
Based on above definition, we have 
\begin{theorem}[Hellmann-Feynman theorem extension] \label{the:4}
The Hellmann-Feynman theorem extension for non-stationary clamped-nuclear-charge alchemical dynamics is
        \begin{align}
            \nabla_Z E & = \nabla_Z \left( \Bra{\Psi_{ns}} \hat{H}_{clamped} \Ket{\Psi_{ns}} \right)\nonumber \\
                       & = \sum_{n,j} c^*_n c_j \exp{\left(\frac{i}{\hbar}\int_0^T \Big(E_n(Z(t'))-E_j(Z(t')) \Big) d t' \right)} \nonumber \\
                        & \qquad \Bra{\chi_n, \phi_n} \nabla_Z \hat{H}_{clamped} \Ket{\chi_j, \phi_j}  + \sum_n E_n \nabla_Z \left(c^*_n c_n\right)  \label{eq:10} 
        \end{align}
where 
\begin{align}
\begin{gathered}
    \Bra{\chi_n, \phi_n} \nabla_Z \hat{H}_{clamped} \Ket{\chi_n, \phi_n} \\
    =  \sum_A \bigg( \frac{\nabla_Z m_A}{{m_A}^2} \big( \Bra{\chi_n} \Delta_{R_A} \Ket{\chi_n} + \Bra{\phi_n} \Delta_{R_A} \Ket{\phi_n} \big)   \\
     - \Bra{\phi_n}  \sum_i \frac{1}{|r_i-R_A|} \Ket{\phi_n}  +  \Bra{\chi_n}\frac{1}{2} \sum_{B\neq A} \frac{Z_B}{|R_A-R_B|} \Ket{\chi_n} \bigg) \label{eq:34} 
\end{gathered}
\end{align}
$\sum_n E_n \nabla_Z \left(c^*_n c_n\right)$ is the additional term comparing to the conventional Hellmann-Feynman theorem.
\end{theorem}
All the variables in the expressions of wave functions are omitted for the sake of simplicity. The proof of this theorem is in \ref{app:4}. 


\section{Summary} \label{sec:8}
A chemical reaction could be seen as evolving along not only nuclear coordinates but also nuclear charges from the computational point of view. We have introduced the pseudo-mass parameterized alchemical equation, which provides a scheme to equally treat nuclear charges and coordinates. A heavy pseudo-mass of nuclear charge is assumed, the nuclear charge wave function and the molecular wave function are well separated, just like molecular wave function and electronic wave function are decoupled in the Born-Oppenheimer approximation. 
Follow this idea, we have discussed some closely related concepts about the alchemical equation, including its function space, geometric phase, phase space, and a Hellmann-Feynman theorem extension for the non-stationary situation, etc. Some interesting properties come out. There are three terms in the geometric phase change expression, contributions coming from both closed nuclear charge $Z$ path and coordinate $R$ path. Momentum-like operator for the nuclear charge is no longer self-adjoint, phase space concept needs to be generalised. The alchemical equation could also be simplified to a concise form for hydrogen-like atom, a Hamiltonian with strong repulsive inverse square potential.

\section*{Acknowledgement}
The author thanks Xiang-Da Peng for fruitful discussions and
Anatole von Lilienfeld for kind support at the University of Basel.


\appendix
\renewcommand\thesection{Appendix \Alph{section}} 
\section{Proof of Theorem \ref{the:1}} \label{app:1}
\begin{proof}
First we use alchemical operator (\ref{eq:27}) acting on the total wave function (\ref{eq:12}), 
\begin{equation}
    \sum_A \left( -\frac{1}{\mu} \Delta_{Z_A} - \frac{1}{m_A} \Delta_{R_{A}} + H_e + \frac{1}{2}\sum_{B \neq A} \frac{Z_A Z_B}{|R_A-R_B|} \right) \eta_n \chi_n \phi_n = E_n \eta_n \chi_n \phi_n \label{eq:36}
\end{equation}
in which 
\begin{align}
    \Delta_{Z_A} \left(\eta_n \chi_n \phi_n \right)& = \nabla_{Z_A}\big( \nabla_{Z_A} \left(\eta_n \chi_n \phi_n \right) \big) \nonumber \\
    & =  \chi_n \phi_n \Delta_{Z_A} \eta_n + \eta_n \phi_n \Delta_{Z_A} \chi_n + \eta_n \chi_n \Delta_{Z_A} \phi_n  \nonumber \\
    & \quad + 2 \phi_n \left( \nabla_{Z_A} \eta_n \right) \left( \nabla_{Z_A} \chi_n \right) + 2 \chi_n \left( \nabla_{Z_A} \eta_n \right) \left( \nabla_{Z_A} \phi_n \right) \nonumber \\
    & \quad + 2 \eta_n \left( \nabla_{Z_A} \chi_n \right) \left( \nabla_{Z_A} \phi_n \right) \label{eq:37}  \\
    \Delta_{R_A} \left( \eta_n \chi_n \phi_n \right) & = \eta_n \nabla_{R_A} \big( \nabla_{R_A} \left( \chi_n \phi_n \right) \big) \nonumber \\
    & = \eta_n \big( \phi_n \Delta_{R_A} \chi_n + 2 \left( \nabla_{R_A} \chi_n \right) \left( \nabla_{R_A} \phi_n \right)  + \chi_n \Delta_{R_A} \phi_n \big) \label{eq:38}
\end{align}
Then multiplying $\chi_n^*$ and $\phi_n^*$ on both sides of Eq.~(\ref{eq:36}) from left and integrate.
Based on conditions (\ref{eq:26}), we have
\begin{align}
    \Bra{\chi_n, \phi_n}\sum_A \left( -\frac{1}{\mu} \Delta_{Z_A} - \frac{1}{m_A} \Delta_{R_{A}} + H_e + \frac{1}{2}\sum_{B \neq A} \frac{Z_A Z_B}{|R_A-R_B|} \right) \Ket{\eta_n, \chi_n, \phi_n} = E_n \eta_n \label{eq:42}
\end{align}
and
\begin{align}
    \Bra{\chi_n, \phi_n}\Delta_{Z_A} \Ket{\eta_n, \chi_n, \phi_n} & = \left( \Delta_{Z_A}  + \Bra{\chi_n} \Delta_{Z_A} \Ket{\chi_n} + \Bra{\phi_n} \Delta_{Z_A} \Ket{\phi_n} \right) \eta_n \label{eq:39} \\
    \Bra{\chi_n, \phi_n} \Delta_{R_A} \Ket{\eta_n, \chi_n, \phi_n} & = \left( \Bra{\chi_n} \Delta_{R_A} \Ket{\chi_n} + \Bra{\phi_n} \Delta_{R_A} \Ket{\phi_n} \right) \eta_n \label{eq:30}
\end{align}
Note that $\Bra{v}\nabla_w\Ket{v} = 0$ where $v=\chi_n$ or $\phi_n$, $w={Z_A}$ or ${R_A}$ when the wave functions can be made real. 
Coupling terms which involve two different $n$ are also vanished, i.e. $\Bra{u}\nabla_w\Ket{v} = 0$ where $v=\chi_n$ or $\phi_n$, $u=\chi_m$ or $\phi_m$, $n \neq m$, $w={Z_A}$ or ${R_A}$. 
Moreover, based on the Born-Oppenheimer approximation and assumption $\mu \gg m $, coupling terms between $r$ and $R, Z$ could be treated separately, so does $R$ and $Z$. Then we have 
\begin{equation}
    \Bra{\chi_n, \phi_n} H_e \Ket{\eta_n, \chi_n, \phi_n}  = \Bra{\phi_n} H_e \Ket{\phi_n} \eta_n \label{eq:40}
\end{equation}
\begin{equation}
    \Bra{\chi_n, \phi_n} \frac{1}{2}\sum_{B \neq A} \frac{Z_A Z_B}{|R_A-R_B|} \Ket{\eta_n, \chi_n, \phi_n}  = \Bra{\chi_n}\frac{1}{2}\sum_{B \neq A}  \frac{Z_A Z_B}{|R_A-R_B|} \Ket{\chi_n} \eta_n \label{eq:41} 
\end{equation}
In the end we substitute Eq.~(\ref{eq:39}), (\ref{eq:30}), (\ref{eq:40}) and (\ref{eq:41}) into (\ref{eq:42}), omit subscript $n$ and finally get Theorem \ref{the:1}.
\end{proof}


\section{Proof of Theorem \ref{the:2}}  \label{app:2}
\begin{proof}
This proof follows similar routines as in Berry phase's derivation \cite{Bohm1992}.
Since alchemical operator (\ref{eq:27}) does not commute at different times, 
i.e. Eq.~(\ref{eq:7}) exist, electronic coordinate $r$ does not evolve with time, we have 
\begin{equation}
    \Psi_n(t) = c_n(t)~ \eta_n(Z(t))~ \chi_n(R(t)|Z(t))~\phi_n(r | (R(t), Z(t))) \label{eq:43}
\end{equation}
Substituting Eq.~(\ref{eq:43}) into time-dependent equation
\begin{equation}
    i \hbar \nabla_t~ \Psi_n(t) = \hat{H}(t)~ \Psi_n(t) \label{eq:44}
\end{equation}
where $\hat{H}(t)$ is the alchemical operator (\ref{eq:27}), we have
\begin{align*}
       & i \hbar \nabla_t  \Big( c_n(t)~ \eta_n(Z(t))~ \chi_n(R(t)|Z(t))~ \phi_n(r | (R(t), Z(t))) \Big) \\
       = & i \hbar ~\phi_n(r | (R(t), Z(t))) \Big( \eta_n(Z(t))~ \chi_n(R(t)|Z(t))~ \nabla_t~ c_n(t)  + \\
       & \qquad c_n(t)~ \chi_n(R(t)|Z(t))~ \nabla_t~ \eta_n(Z(t)) + c_n(t)~ \eta_n(Z(t))~ \nabla_t~ \chi_n(R(t)|Z(t))\Big)   \\
      = & E_n\big(Z(t),R(t)\big)~ c_n(t)~ \eta_n(Z(t))~ \chi_n(R(t)|Z(t))~ \phi_n(r | (R(t), Z(t)))
\end{align*}
Then moving terms and eliminate $\phi_n(r | (R(t), Z(t)))$
\begin{align}
      \eta_n(Z(t))~ \chi_n(R(t)|Z(t))~  \nabla_t~ c_n(t) & = -c_n(t)~ \chi_n(R(t)|Z(t))~ \nabla_t~ \eta_n(Z(t)) \nonumber \\
      & \quad - c_n(t)~ \eta_n(Z(t))~ \nabla_t~ \chi_n(R(t)|Z(t)) \nonumber \\
      & \quad -\frac{i}{\hbar} E_n\big(Z(t),R(t)\big)~ c_n(t)~ \eta_n(Z(t))~ \chi_n(R(t)|Z(t))  \label{eq:45}
\end{align}
Next we multiply $\eta_n^*$ and $\chi_n^*$ on both sides of Eq.~(\ref{eq:45}) from left and integrate.
Based on conditions (\ref{eq:26}), we have
\begin{align*}
    \nabla_t~ c_n(t) & = -c_n(t) \Bra{\eta_n(Z(t))} \nabla_t \Ket{\eta_n(Z(t))} 
    -c_n(t) \Bra{\chi_n(R(t)|Z(t))} \nabla_t \Ket{\chi_n(R(t)|Z(t))} \nonumber \\
    & \quad -\frac{i}{\hbar} E_n\big(Z(t),R(t)\big)~ c_n(t)
\end{align*}
Note that coupling terms which involve two different $n$ are vanished, i.e. $\Bra{v}\nabla_t\Ket{u} = 0$ where $v=\eta_n(Z(t))$ or $\chi_n(R(t)|Z(t))$, $u=\eta_m(Z(t))$ or $\chi_m(R(t)|Z(t))$, $n \neq m$. And further,
\begin{align}
     c_n(t) & = \exp{\left( - \int_0^T \Bra{\eta_n(Z(t'))} \nabla_{t'} \Ket{\eta_n(Z(t'))}~d t'\right)}       \nonumber\\
        & \quad \exp{\left( - \int_0^T \Bra{\chi_n(R(t')|Z(t'))} \nabla_{t'} \Ket{\chi_n(R(t')|Z(t'))}~d t'\right)} \nonumber \\
        & \quad \exp{\left( - \frac{i}{\hbar} \int_0^T E_n(Z(t'),R(t'))~d t'\right)} \label{eq:46}
\end{align}
Substituting Eq.~(\ref{eq:46}) into Eq.~(\ref{eq:43}) and compare with Eq.~(\ref{eq:65}), we have
\begin{equation}
    c_n(C) = \exp{\left(i \gamma_n(C)\right)} \exp{\left( - \frac{i}{\hbar} \int_0^T E_n(Z(t'),R(t'))~d t' \right)}
\end{equation}
Since $\nabla_t = \nabla_Z \frac{\partial Z}{\partial t}$ and $\nabla_t = \nabla_R \frac{\partial R}{\partial t}$, we finally obtain
\begin{align}
     \gamma_n(C) & =   i \oint_{C_Z} \Bra{\eta_n(Z(t))} \nabla_{Z} \Ket{\eta_n(Z(t))}~d Z       \nonumber\\
        & \quad   i \oint_{C_Z} \Bra{\chi_n(R(t)|Z(t))} \nabla_{Z} \Ket{\chi_n(R(t)|Z(t))}~d Z  \nonumber \\
        & \quad   i \oint_{C_R} \Bra{\chi_n(R(t)|Z(t))} \nabla_{R} \Ket{\chi_n(R(t)|Z(t))}~d R  \label{eq:47}
\end{align}
Theorem \ref{the:2} is proved.
\end{proof}


\section{Proof of Theorem \ref{the:3}} \label{app:3}
\begin{proof}
Terms involving $\Delta_R$, $R$ and $\chi$ in the alchemical equation (\ref{eq:1}) could be removed since there is only one atom in the system. Hence we have alchemical equation for hydrogen-like atom
\begin{equation}
    \left( -\frac{1}{\mu} \big(\Delta_{Z} + \Bra{\phi} \Delta_{Z} \Ket{\phi} \big)
    +\Bra{\phi} H_e \Ket{\phi} \right) \eta =E \eta  \label{eq:60} 
\end{equation}
The ground state hydrogen-like wave function with nuclear charge $Z$ is \cite{Levine2014}
\begin{equation}
    \phi_{1s}(r|Z) = \frac{1}{\sqrt{\pi}} \left(\frac{Z}{a}\right)^{\frac{3}{2}}\exp{\left(-\frac{Z r}{a}\right)} \label{eq:61}
\end{equation}
where $a$ is the Bohr radius. Its energy is
\begin{equation}
    \Bra{\phi} H_e \Ket{\phi}  = -Z^2 Ry \label{eq:62}
\end{equation}
Moreover, 
\begin{align}
    \Bra{\phi} \Delta_{Z} \Ket{\phi} & = \Bra{\phi_{1s}(r|Z)} \Delta_{Z} \Ket{\phi_{1s}(r|Z)} \nonumber \\
    & = \int_0^{+\infty} \left( \frac{1}{\sqrt{\pi}}  \left(\frac{Z}{a}\right)^{\frac{3}{2}}\exp{\left(-\frac{Z r}{a}\right)} \right) \Delta_{Z} \left( \frac{1}{\sqrt{\pi}} \left(\frac{Z}{a}\right)^{\frac{3}{2}}\exp{\left(-\frac{Z r}{a}\right)} \right) d r \nonumber \\
    & = -\frac{0.75}{Z^2}\label{eq:63}
\end{align}
Substituting Eq.~(\ref{eq:62}) and Eq.~(\ref{eq:63}) into Eq.~(\ref{eq:60}), we obtain
\begin{equation}
    \left( -\frac{1}{\mu} \left(\Delta_{Z} -\frac{0.75}{Z^2} \right)
     -Z^2 Ry \right) \eta =E \eta  \label{eq:64} 
\end{equation}
In order to bypass the unknown pseudo-mass $\mu$, we use the pseudo-mass weighted nuclear charge, $Z_w = \sqrt{\mu} Z$, to instead $Z$.
$E_w$ and $\eta_w$ are corresponding pseudo-mass weighted energy and the pseudo-mass weighted nuclear charge wave function, respectively.
\begin{equation*}
    \left( -\Delta_{Z_w} + \frac{0.75}{Z_w^2} -\frac{Z_w^2 Ry}{\mu} \right) \eta_w = E_w \eta_w 
\end{equation*}
This is exactly Eq.~(\ref{eq:2}). The last
Eq.~(\ref{eq:3}) is obvious by taking $\mu \to +\infty$ which induces  $-\frac{Z_w^2 Ry}{\mu} \to 0$.
\end{proof}


\section{Proof of Theorem \ref{the:4}} \label{app:4}
\begin{proof}
The force applying on the atom along nuclear charge $Z$ (regardless of its direction) is
\begin{align}
    \nabla_Z E & = \nabla_Z  \Bra{\Psi_{ns}} \hat{H}_{clamped} \Ket{\Psi_{ns}} \nonumber \\
               & =  \Bra{\nabla_Z \Psi_{ns}} \hat{H}_{clamped} \Ket{\Psi_{ns}}  +  \Bra{\Psi_{ns}} \nabla_Z \hat{H}_{clamped} \Ket{\Psi_{ns}}  +  \Bra{\Psi_{ns}} \hat{H}_{clamped} \Ket{\nabla_Z \Psi_{ns}}  \label{eq:49} 
\end{align}
in which 
\begin{align}
     \Bra{\Psi_{ns}} \nabla_Z \hat{H}_{clamped} \Ket{\Psi_{ns}}  &  = \sum_{n,j} c^*_n c_j \exp{\left(\frac{i}{\hbar}\int_0^T \Big(E_n(Z(t'))-E_j(Z(t')) \Big) d t' \right)} \nonumber \\
    & \qquad \Bra{\chi_n, \phi_n} \nabla_Z \hat{H}_{clamped} \Ket{\chi_j, \phi_j} \label{eq:50}
\end{align}
and
\begin{align}
\begin{gathered}
    \Bra{\chi_n, \phi_n} \nabla_Z \hat{H}_{clamped} \Ket{\chi_j, \phi_j} \\
    = \Bra{\chi_n, \phi_n} \sum_A \bigg( \frac{\nabla_Z m_A}{m_A^2} \Delta_{R_{A}} - \sum_i \frac{1}{|r_i-R_A|}   + \frac{1}{2} \sum_{B \neq A} \frac{Z_B}{|R_A-R_B|} \bigg) \Ket{\chi_j, \phi_j} \\ 
      = \sum_A \bigg( \frac{\nabla_Z m_A}{m_A^2}  \Big( \Bra{\chi_n}\Delta_{R_{A}} \Ket{\chi_n} + \Bra{\phi_n}\Delta_{R_{A}} \Ket{\phi_n} \Big)   \\ 
      - \sum_i \Bra{\phi_n}\frac{1}{|r_i-R_A|}\Ket{\phi_n}  + \Bra{\chi_n}\frac{1}{2} \sum_{B \neq A} \frac{Z_B}{|R_A-R_B|} \Ket{\chi_n} \bigg) \label{eq:59}
\end{gathered}
\end{align}
This is exactly Eq.~(\ref{eq:34}).
In order to calculate the other two terms in Eq.~(\ref{eq:49}), let us first consider $\nabla_Z \Psi_{ns}$. 
\begin{align}
    \nabla_Z \Psi_{ns} & = \sum_j c_j \exp{\left( -\frac{i}{\hbar}\int_0^T E_j\left(Z(t')\right) d t' \right)} \Biggl( \chi_j \nabla_Z \phi_j + \phi_j \nabla_Z \chi_j \nonumber \\
    & \qquad -\frac{i}{\hbar} \chi_j \phi_j \nabla_Z \left( \int_0^T  E_j\left(Z(t')\right) d t' \right) 
    \Biggr) + \sum_j \left( \nabla_Z c_j \right) \nonumber \\
    & \qquad \exp{\left( -\frac{i}{\hbar}\int_0^T E_j\left(Z(t')\right) d t' \right)} \chi_j \phi_j \label{eq:51}
\end{align}
Next we apply $\hat{H}_{clamped}$ acting on Eq.~(\ref{eq:51}) and multiply $\chi_n^*$ and $\phi_n^*$ on both sides from left and integrate. Based on conditions (\ref{eq:26}), we have
\begin{align}
\begin{gathered}
    \Bra{\Psi_{ns} } \hat{H}_{clamped} \Ket{ \nabla_Z \Psi_{ns} } \\
    = \sum_{n,j} c_n^* c_j \exp{\left(\frac{i}{\hbar}\int_0^T \Big(E_n(Z(t'))-E_j(Z(t'))\Big) d t'  \right)}  \\
     \Biggl( \Bra{\chi_n, \phi_n } \hat{H}_{clamped} \Ket{ \chi_j, \nabla_Z  \phi_j } 
    +   \Bra{\chi_n, \phi_n } \hat{H}_{clamped} \Ket{ \nabla_Z  \chi_j, \phi_j  }   \\
    -\frac{i}{\hbar} \sum_n c^2_n E_n \nabla_Z \left( \int_0^T  E_n\left(Z(t')\right) d t' \right) \Biggr) + \sum_n c^*_n E_n \nabla_Z c_n  \label{eq:52}
\end{gathered}
\end{align}
Similarly, we have
\begin{align}
\begin{gathered}
    \Bra{\nabla_Z  \Psi_{ns} } \hat{H}_{clamped} \Ket{\Psi_{ns}} \\
    = \sum_{j,n} c_j^* c_n \exp{\left(\frac{i}{\hbar}\int_0^T \Big(E_j(Z(t'))-E_n(Z(t'))\Big) d t'  \right)}   \\
    \Biggl( \Bra{\chi_j, \nabla_Z  \phi_j} \hat{H}_{clamped} \Ket{\chi_n, \phi_n } 
    +   \Bra{\nabla_Z  \chi_j, \phi_j} \hat{H}_{clamped} \Ket{\chi_n, \phi_n }    \\
     +\frac{i}{\hbar} \sum_n c^2_n E_n \nabla_Z \left( \int_0^T  E_n\left(Z(t')\right) d t' \right) \Biggr) + \sum_n c_n E_n \nabla_Z c_n^*   \label{eq:53}
\end{gathered}
\end{align}
Summing Eq.~(\ref{eq:52}) and Eq.~(\ref{eq:53}) up, we get 
\begin{align}
\begin{gathered}
    \Bra{\Psi_{ns} } \hat{H}_{clamped} \Ket{ \nabla_Z \Psi_{ns} }+ \Bra{\nabla_Z  \Psi_{ns} } \hat{H}_{clamped} \Ket{\Psi_{ns}} \\
    =  \sum_{j,n} c_j^* c_n \exp{\left(\frac{i}{\hbar}\int_0^T \left(E_j(Z(t'))-E_n(Z(t')) \right) d t'  \right)}   \\
      \Big( E_n \Braket{\chi_j, \nabla_Z  \phi_j | \chi_n, \phi_n } + E_n \Braket{\nabla_Z  \chi_j, \phi_j | \chi_n, \phi_n }  \\
      +  E_j \Braket{\chi_j, \phi_j | \chi_n, \nabla_Z \phi_n  } + E_j \Braket{\chi_j, \phi_j |   \nabla_Z \chi_n, \phi_n } \Big)   \\
      + \sum_n E_n \nabla_Z \left(  c^*_n c_n \right)  \label{eq:54} 
\end{gathered}
\end{align}
Since 
\begin{align}
     \Braket{\chi_j, \nabla_Z  \phi_j | \chi_n, \phi_n }  & = - \Braket{\chi_j, \phi_j | \chi_n, \nabla_Z \phi_n } \nonumber \\
     \Braket{\nabla_Z  \chi_j, \phi_j | \chi_n, \phi_n }  & = - \Braket{\chi_j, \phi_j | \nabla_Z \chi_n, \phi_n } \label{eq:55}
\end{align}
Eq.~(\ref{eq:54}) turns to
\begin{align}
\begin{gathered}
    \Bra{\Psi_{ns} } \hat{H}_{clamped} \Ket{ \nabla_Z \Psi_{ns} }+ \Bra{\nabla_Z  \Psi_{ns} } \hat{H}_{clamped} \Ket{\Psi_{ns}}   \\
    =  \sum_{j,n} c_j^* c_n \exp{\left(\frac{i}{\hbar}\int_0^T \left(E_j(Z(t'))-E_n(Z(t'))\right) d t'  \right)}    \\
     \Big( \left(E_j - E_n \right) \Braket{\chi_j, \phi_j | \chi_n, \nabla_Z \phi_n  }  + \left(E_j - E_n \right) \Braket{\chi_j, \phi_j | \nabla_Z \chi_n, \phi_n  } \Big)    \\
     + \sum_n E_n \nabla_Z \left(  c^*_n c_n \right)   \\
     =  \sum_{j,n} c_j^* c_n \exp{\left(\frac{i}{\hbar}\int_0^T \left(E_j(Z(t'))-E_n(Z(t'))\right) d t'  \right)}   \\
     \Big( \left(E_j - E_n \right) \delta_{j,n} \Braket{\phi_j | \nabla_Z \phi_n  }   
      + \left(E_j - E_n \right) \delta_{j,n} \Braket{\chi_j| \nabla_Z \chi_n } \Big)  \\
      + \sum_n E_n \nabla_Z \left(  c^*_n c_n \right)     \\
     = \sum_n E_n \nabla_Z \left(  c^*_n c_n \right) \label{eq:56} 
\end{gathered}
\end{align}
Combining Eq.~(\ref{eq:56}) and Eq.~(\ref{eq:50}), we finally obtain Eq.~(\ref{eq:10}).
\end{proof}

\bibliographystyle{unsrt}

\begin{thebibliography}{10}

\bibitem{VonLilienfeld2005}
O.~Anatole von Lilienfeld, Roberto~D. Lins, and Ursula Rothlisberger.
\newblock {Variational Particle Number Approach for Rational Compound Design}.
\newblock {\em Physical Review Letters}, 95(15):153002, 2005.

\bibitem{Balawender2019}
Robert Balawender, Michael Lesiuk, Frank {De Proft}, Christian {Van Alsenoy},
  and Paul Geerlings.
\newblock {Exploring chemical space with alchemical derivatives: alchemical
  transformations of H through Ar and their ions as a proof of concept}.
\newblock {\em Physical Chemistry Chemical Physics}, 21(43):23865--23879, 2019.

\bibitem{Chang2014}
K.~Y.Samuel Chang and O.~Anatole {Von Lilienfeld}.
\newblock {Quantum mechanical treatment of variable molecular composition: From
  `alchemical' changes of state functions to rational compound design}.
\newblock {\em Chimia}, 68(9):602--608, 2014.

\bibitem{Hahn2019}
David~F. Hahn and Philippe~H. H{\"{u}}nenberger.
\newblock {Alchemical Free-Energy Calculations by Multiple-Replica
  $\lambda$-Dynamics: The Conveyor Belt Thermodynamic Integration Scheme}.
\newblock {\em Journal of Chemical Theory and Computation}, 15(4):2392--2419,
  2019.

\bibitem{Marx2009}
Dominik Marx and Jurg Hutter.
\newblock {\em {Ab Initio Molecular Dynamics}}.
\newblock Cambridge University Press, Cambridge, 2009.

\bibitem{Gatti2014}
Fabien Gatti, editor.
\newblock {\em {Molecular Quantum Dynamics}}.
\newblock Physical Chemistry in Action. Springer Berlin Heidelberg, Berlin,
  Heidelberg, 2014.

\bibitem{Bonneau2001}
Guy Bonneau, Jacques Faraut, and Galliano Valent.
\newblock {Self-adjoint extensions of operators and the teaching of quantum
  mechanics}.
\newblock {\em American Journal of Physics}, 69(3):322--331, 2001.

\bibitem{Araujo2004}
Vanilse~S. Araujo, F.~A.~B. Coutinho, and J.~{Fernando Perez}.
\newblock {Operator domains and self-adjoint operators}.
\newblock {\em American Journal of Physics}, 72(2):203--213, 2004.

\bibitem{Scursulim2020}
J~V~S Scursulim, A~A Lima, U.~Camara da~Silva, and G~M Sotkov.
\newblock {Supersymmetry shielding the scaling symmetry of conformal quantum
  mechanics}.
\newblock {\em Physical Review A}, 101(3):032105, 2020.

\bibitem{BERRY1984}
M.~V. Berry.
\newblock {Quantal phase factors accompanying adiabatic changes}.
\newblock {\em Proceedings of the Royal Society of London. A. Mathematical and
  Physical Sciences}, 392(1802):45--57, 1984.

\bibitem{Mead1992}
C.~Alden Mead.
\newblock {The geometric phase in molecular systems}.
\newblock {\em Reviews of Modern Physics}, 64(1):51--85, 1992.

\bibitem{Bohm1992}
A.~Bohm, B.~Kendrick, and Mark~E. Loewe.
\newblock {The Berry phase in molecular physics}.
\newblock {\em International Journal of Quantum Chemistry}, 41(1):53--75, 1992.

\bibitem{Twamley2006}
J.~Twamley and G.~J. Milburn.
\newblock {The quantum Mellin transform}.
\newblock {\em New Journal of Physics}, 8(12):328--328, 2006.

\bibitem{Coon2002}
Sidney~A. Coon and Barry~R. Holstein.
\newblock {Anomalies in quantum mechanics: The $1/r^2$ potential}.
\newblock {\em American Journal of Physics}, 70(5):513--519, 2002.

\bibitem{Essin2006}
Andrew~M. Essin and David~J. Griffiths.
\newblock {Quantum mechanics of the $1/x^2$ potential}.
\newblock {\em American Journal of Physics}, 74(2):109--117, 2006.

\bibitem{Moroz2010}
Sergej Moroz and Richard Schmidt.
\newblock {Nonrelativistic inverse square potential, scale anomaly, and complex
  extension}.
\newblock {\em Annals of Physics}, 325(2):491--513, 2010.

\bibitem{Nishida2007}
Yusuke Nishida and Dam~T. Son.
\newblock {Nonrelativistic conformal field theories}.
\newblock {\em Physical Review D}, 76(8):086004, 2007.

\bibitem{Camblong2000}
Horacio~E. Camblong, Luis~N. Epele, Huner Fanchiotti, and Carlos~A.
  {Garc{\'{i}}a Canal}.
\newblock {Renormalization of the Inverse Square Potential}.
\newblock {\em Physical Review Letters}, 85(8):1590--1593, 2000.

\bibitem{Bala1994}
P.~Bala, B~Lesyng, and J.A. McCammon.
\newblock {Extended Hellmann-Feynman theorem for non-stationary states and its
  application in quantum-classical molecular dynamics simulations}.
\newblock {\em Chemical Physics Letters}, 219(3-4):259--266, 1994.

\bibitem{Levine2014}
Ira~N. Levine.
\newblock {\em {Quantum Chemistry}}.
\newblock Pearson, 7th edition, 2014.

\end{thebibliography}

\end{document}